\PassOptionsToPackage{bookmarks=false}{hyperref}
\documentclass[sigchi]{acmart}
\acmConference{}{}{}
\renewcommand\footnotetextcopyrightpermission[1]{} 
\usepackage{hyphenat}
\usepackage{booktabs} 
\usepackage{tabularx}
\usepackage{textcomp}
\usepackage{graphicx}
\usepackage{caption}
\usepackage{subcaption}
\usepackage{amssymb}
\usepackage{amsmath,amssymb,amsfonts}
\usepackage{algorithmic}
\usepackage{textcomp}
\usepackage{multirow}
\usepackage{xcolor}
\usepackage{adjustbox}
\usepackage{caption}

\def\BibTeX{{\rm B\kern-.05em{\sc i\kern-.025em b}\kern-.08emT\kern-.1667em\lower.7ex\hbox{E}\kern-.125emX}}
\begin{document}
%
\title {HTMLPhish: Enabling Phishing Web Page Detection by Applying Deep Learning Techniques on HTML Analysis}

\author{Chidimma Opara}
\affiliation{%
  \institution{Teesside University}
  \city{Middlesbrough}
  \postcode{TS1 3BX}
  \country{UK}}
\email{c.opara@tees.ac.uk}
\author{Bo Wei}
\affiliation{%
  \institution{Northumbria University}
  \city{Newcastle upon Tyne}
  \country{UK}
}
\email{bo.wei@northumbria.ac.uk}
\author{Yingke Chen}
\affiliation{%
 \institution{Teesside University}
 \city{Middlesbrough}
 \country{UK}}
\email{y.chen@tees.ac.uk}

\fancyfoot{}

\begin{abstract}
Recently, the development and implementation of phishing attacks require little technical skills and costs. This uprising has led to an ever-growing number of phishing attacks on the World Wide Web. Consequently, proactive techniques to fight phishing attacks have become extremely necessary. In this paper, we propose HTMLPhish, a deep learning based data-driven end-to-end automatic phishing web page classification approach. Specifically, HTMLPhish receives the content of the HTML document of a web page and employs Convolutional Neural Networks (CNNs) to learn the semantic dependencies in the textual contents of the HTML. The CNNs learn appropriate feature representations from the HTML document embeddings without extensive manual feature engineering. Furthermore, our proposed approach of the concatenation of the word and character embeddings allows our model to manage new features and ensure easy extrapolation to test data. We conduct comprehensive experiments on a dataset of more than 50,000 HTML documents that provides a distribution of phishing to benign web pages obtainable in the real-world that yields over 93\% Accuracy and True Positive Rate. Also, HTMLPhish is a completely language-independent and client-side strategy which can, therefore, conduct web page phishing detection regardless of the textual language.
\end{abstract}


\maketitle

\pagestyle{plain}

\section{Introduction}
The infamous phishing attack is a social engineering technique that manipulates internet users into revealing private information that may be exploited for fraudulent purposes \cite {lopez2018access}. This form of cybercrime has recently become common because it is carried out with little technical ability and significant cost \cite {sahingoz2019machine}. The proliferation of phishing attacks is evident in the 46\% increase in the number of phishing websites identified between October 2018 and March 2019 by the Anti-Phishing Working Group (APWG) \cite{APWGReports}. Most phishing attacks are started by an unsuspecting Internet user merely clicking on a link in a phishing email message that leads to a bogus website. The impact of phishing attacks on individuals such as identity theft, psychological, and financial costs can be devastating.

\subsection{Problem Definition}
Recent research in phishing detection approaches has resulted in the rise of multiple technical methods such as augmenting password logins \cite{chattaraj2018new}, and multi-factor authentication \cite{acar2013single}. However, these techniques are usually server-side systems that require the Internet user to correspond with a remote service, which adds further delay in the communication channel. Another popular phishing detection system that relies on a centralised architecture is the phishing blacklist and whitelist methods \cite{Google}. A URL visited by an internet user will be compared with the URL in these lists in real-time. Although the list based methods tend to keep the false positive rate low, however, a significant shortcoming is that the lists are not exhaustive, and they fail to detect zero-day phishing attacks. To mitigate these limitations, researchers have developed several anti-phishing techniques using machine learning models as they are mostly client-side based and can generalise their predictions on unseen data.

Machine learning-based anti-phishing techniques typically follow specific approaches: (1) The required representation of features is firstly extracted, then (2) a phishing detection machine learning model is trained using the feature vectors. To extract the feature representation from the lexical and static components of a web page, the machine learning models rely on the assumption that the infrastructure of phishing pages are different from legitimate pages. For example, in \cite {amrutkar2017detecting}, phishing web pages are automatically detected based on handcrafted features extracted from the URL, HTML content, network, and JavaScript of a web page. Furthermore, natural language processing techniques are currently used to extract specific features such as the number of common phishing words, type of ngram, etc. from the components of a web page \cite{lecun2015deep,gutierrez2018learning,buber2017detecting}.

While the above approaches have proven successful, they nevertheless are prone to several limitations, particularly in the context of HTML analysis: \textit{i. inability to accommodate unseen features:} As the accuracy of existing models depends on how comprehensive the feature set is and how impervious the feature set remains to future attacks, they will be unable to correctly detect new phishing web pages with evolved content and structure without a regular update of the feature set. \textit{ii. They require substantial manual feature engineering:} Existing phishing detection machine learning models require specialised domain knowledge in order to ascertain the needed features suitable to each task (e.g., number of white spaces in the HTML content, number of redirects, and iframes, etc.). This is a tedious process, and these handcrafted features are often targeted and bypassed in future attacks. It is also challenging to know the best features for one particular application.

To address the above issues, we propose HTMLPhish, a deep learning based data-driven end-to-end automatic phishing web page classification approach. Specifically, HTMLPhish uses both the character and word embedding techniques to represent the features of each HTML document. Then Convolutional Neural Networks (CNNs) are employed to model the semantic dependencies.

The following characteristics highlight the relevance of HTMLPhish to web page phishing detection: 

(1) HTMLPhish analyses HTML directly to help reserve useful information. It also removes the arduous task required for the manual feature engineering process. 

(2) HTMLPhish takes into consideration all the elements of an HTML document, such as text, hyperlinks, images, tables, and lists, when training the deep neural network model.

We experimentally demonstrate the significance of character and word embedding features of HTML contents in detecting phishing web pages. We then propose a state-of-the-art HTML phishing detection model, in which the character and word embedding matrices are concatenated before employing convolutions on the represented features. Our proposed approach ensures an adequate embedding of new feature vectors that enables straightforward extrapolation of the trained model to test data. Subsequently, we conduct extensive evaluations on a dataset of over 50,000 HTML documents collected over two months. This ensures our evaluation settings reproduces real-world situations in which models are applied to data generated up to the present point and applied to new data. 

We summarise the main contributions of this paper as follows:
\begin{itemize}
\item Different from existing methods, our proposed model, HTMLPhish, to the best of our knowledge, is the first to use only the raw content of the HTML document of a web page to train a deep neural network model for phishing detection. Manual feature engineering is reduced as HTMLPhish learns the representation in the features of the HTML document, and we do not depend on any other complicated or specialist features for the task. Our proposed approach takes advantage of the word and character embedding matrix to present a phishing detection model that automatically accommodates new features and is therefore easily applied to test data. 
\item We conduct extensive evaluations on a dataset of more than 50, 000 HTML documents collected in two months. The distribution of the instances in our dataset is similar to the ratio of phishing and legitimate web pages found in the real-world. This ensures that our evaluation metrics and results are relevant to existing systems. 
\item Furthermore, we carried out a longitudinal study on the efficiency HTMLPhish to infer the maximum retraining period, for which the accuracy of the system does not reduce. Our result only recorded a minimal 4\% decrease in accuracy on the test data. This confirms that HTMLPhish remains reliable and temporally robust over a long period.
\end{itemize}

We organised the remainder of the paper as follows: the next section provides an overview of related works on proposed techniques of detecting phishing on web pages. Section 3 gives the prior knowledge on Convolutional Neural Networks, and Section 4 provides an in-depth description of our proposed model. Section 5 elaborates on the dataset collection, while the detailed results on the evaluations of our proposed model are found in Section 6. Finally, we conclude our paper in Section 7.

\section{Related Works}
In this section, we address two most closely related topics to our work: the phishing web page detection using feature engineering and the Deep Learning method (especially for NLP).
\subsection{Feature Engineering for Phishing Web Page Detection}
These techniques extract specific features from a web page such as JavaScript, HTML web page, URL, and network features. These are fed into machine learning algorithms to build a classification model. These machine learning techniques differ in the type of heuristics and number of feature sets used and the optimisation algorithm applied to the machine learning algorithm. These techniques are based on the fact that both the phishing and benign web pages have a different content distribution of extracted features. The accuracy of heuristics and machine learning-based techniques critically depends on the type of features extracted, and the machine learning algorithm applied. Many phishing detection techniques have been built on different proposed feature sets.

Varshney et al \cite{ varshney2016phish} proposed LPD, a client-side based web page phishing detection mechanism. The strings from the URL and page title from a specified web page is extracted and searched on the Google search engine. If there is a match between the domain names of the top T search results and the domain name of the specified URL, the web page is considered to be legitimate. The result from their evaluations gave a true positive rate of 99.5\%.

Smadi et al. \cite{smadi2018detection} proposed a neural network model that can adapt to the dynamic nature of phishing emails using reinforcement learning. The proposed model can handle zero-day phishing attacks and also mitigate the problem of a limited dataset using an updated offline database. Their experiment yielded a high accuracy of 98.63\% on fifty features extracted from a dataset of 12,266 emails.

The selection of features from various web page elements can be an expensive process from security risk and technological workload angle. For example, it can be prolonged and somewhat problematic to extract specific feature sets. Besides, it needs specialist domain expertise to define which features are essential.

\subsection{Deep Learning}
Due to its performance in many applications, Deep Learning has attracted increased interest in recent years \cite{goodfellow2016deep, he2016deep, Goodfellow-et-al-2016}. The core concept is to learn the feature representation from unprocessed data instantaneously without any manual feature engineering. Under this premise, we want to use Deep Learning to detect phishing HTML content by directly learning how features from the raw HTML string is represented instead of using specialist features that are manually engineered.

As we want to train our Deep Learning networks using textual features, it is, therefore, essential to discuss NLP as it relates to Deep Learning. Deep learning techniques have been successful in a lot of NLP tasks, for example, in document classification \cite{kim2014convolutional}, machine translation \cite{cho2014learning}, etc. Recurrent neural networks (e.g., LSTM \cite{hochreiter1997long}) have been extensively applied due to their ability to exhibit temporal behaviour and capture sequential data. However, CNN has become brilliant substitutes for LSTMs, especially showing excellent performance in text classification and sentiment analysis as CNN learns to recognize patterns across space \cite{yin2017comparative}.

Very few attempts have been made to use Deep Learning to detect phishing web pages using web page components. Bahnsen et al. \cite{bahnsen2017classifying} proposed a phishing classifying scheme that used features of the URLs of a web page as input and implemented the model on an LSTM network. The results yielded gave an accuracy of 98.7\% accuracy on a corpus of 2 million phishing and legitimate URLs. The authors of \cite{le2018urlnet} proposed a CNN based model which combines the outputs of two Convolutional layers to detect malicious URLs. 

However, our review did not find any existing approach that detects malicious phishing web pages using only HTML documents on Deep Learning. HTMLPhish learns the semantic information present only in the character and words in an HTML document to determine the maliciousness of the web page. Our thorough analysis shows that phishing web pages can be detected using only their HTML document content.

\section{Preliminaries}

We define the problem of detecting phishing web pages using their HTML content as a binary classification task for prediction of two classes: \textit{legitimate} or \textit{phishing}. Given a dataset with \textit{T} HTML documents
${\{(html_1, y_1), . . . , (html_\textit{T}, y_\textit{T})}\}$, where ${html_t}$ for \textit{t} = 1, . . . , \textit{T} represents an HTML document , while ${y_t\in \{0, 1\}}$ is its label. ${{y_t = 1}}$ corresponds to a phishing HTML document while ${{y_t = 0}}$ is a legitimate HTML document.

\subsection{Deep Neural Network for Phishing HTML Document Detection}
The deep neural network that underlies HTMLPhish is a Convolutional Neural Network (CNN). To detail a basic CNN for HTML document classification, an HTML document is comprised of a string of characters or words. Our goal is to obtain an embedding matrix \textit{html} \textrightarrow s $ \epsilon \mathbb{R}^{maxlen \times d} $, in a way that \textit{s} is made up of sets of adjoining inputs $s_i \in (1, 2,..., maxlen)$ in a string, in which the input can be individual characters or words from the HTML document. Each input is subsequently transformed in an embedding $s_{i} \epsilon \mathbb{R}^{d}$ is the $i^{th}$ column of \textit{S} and the \textit{d-dimension} is the vector size which is automatically initialized and learnt together with the remainder of the model.

In this paper, the embedding matrix was automatically initialised, and for parallelisation, all sequences were padded to the same length \textit{maxlen}.

The CNN performs a convolution operation $\otimes$ over $ s \epsilon \mathbb{R}^{maxlen \times d} $ using: $$ c_i = f(M \otimes s_{i:i+n-1} + b_i) $$ followed by a non-linear activation where $b_i$ is the bias, \textit{M} is the convolving filter and \textit{n} is the kernel size of the convolution operation. After the convolution, a pooling step is applied (which in our model is the Max Pooling) in order to decrease the feature dimension and determine the most important features.

The CNN is capable of exploiting the temporal relation of \textit{n} kernel size in its input using the filter \textit{M} to convolve on each segment of \textit{n} kernel size. A CNN model typically contains several sets of filters with different kernel sizes \textit{(n)}. Those are the model hyperparameters that are set by the user. In this deep neural network, the convolution layer is usually followed by a Pooling layer. The features from the Pooling layer are then passed to dense layers to perform the required classification. The entire network is then trained by using backpropagation.

\textbf{Note:} In order to differentiate our state-of-the-art model from the baseline models, for the rest of this paper, we will use the term HTMLPhish-Full to indicate HTMLPhish trained with the proposed model unless otherwise stated, while HTMLPhish-Character and HTMLPhish-Word represent the deep neural network model using only the character and word embedding respectively.

\section {The Proposed Model}
In this section, we elaborate on the architecture of our proposed deep neural network model HTMLPhish-Full. The network architecture seen in Figure \ref{fig:HTMLPhish} shows HTMLPhish-Full has two input layers. The first input layer processes the raw HTML document into an embedding matrix made up of character-level feature representations, while the second input layer does the same with words. These two branches are concatenated in a dense layer called the Concatenation layer. Therefore, the embedding matrix in this model is the sum of the character-level embedding matrix and the word embedding matrix  $C_{em} $ + $W_{em}$ where $C_{em} $ \textrightarrow c $ \epsilon \mathbb{R}^{maxlen_1 \times d}$, and $W_{em}$ \textrightarrow w $ \epsilon \mathbb{R}^{maxlen_2 \times d}$. The features in the Concatenation layer allows the preservation of the original information in the HTML content. In the concatenation layer, the content of both embedding layers are put alongside each other to yield a 3 dimensional layer [$C_{em} $ + $W_{em}$ \textrightarrow (None, 180, 100) + (None, 2000, 100)  = (None, 2180, 100)].

\begin{figure}[ht]
\centering
\includegraphics[width = \columnwidth, height = 4.5cm]{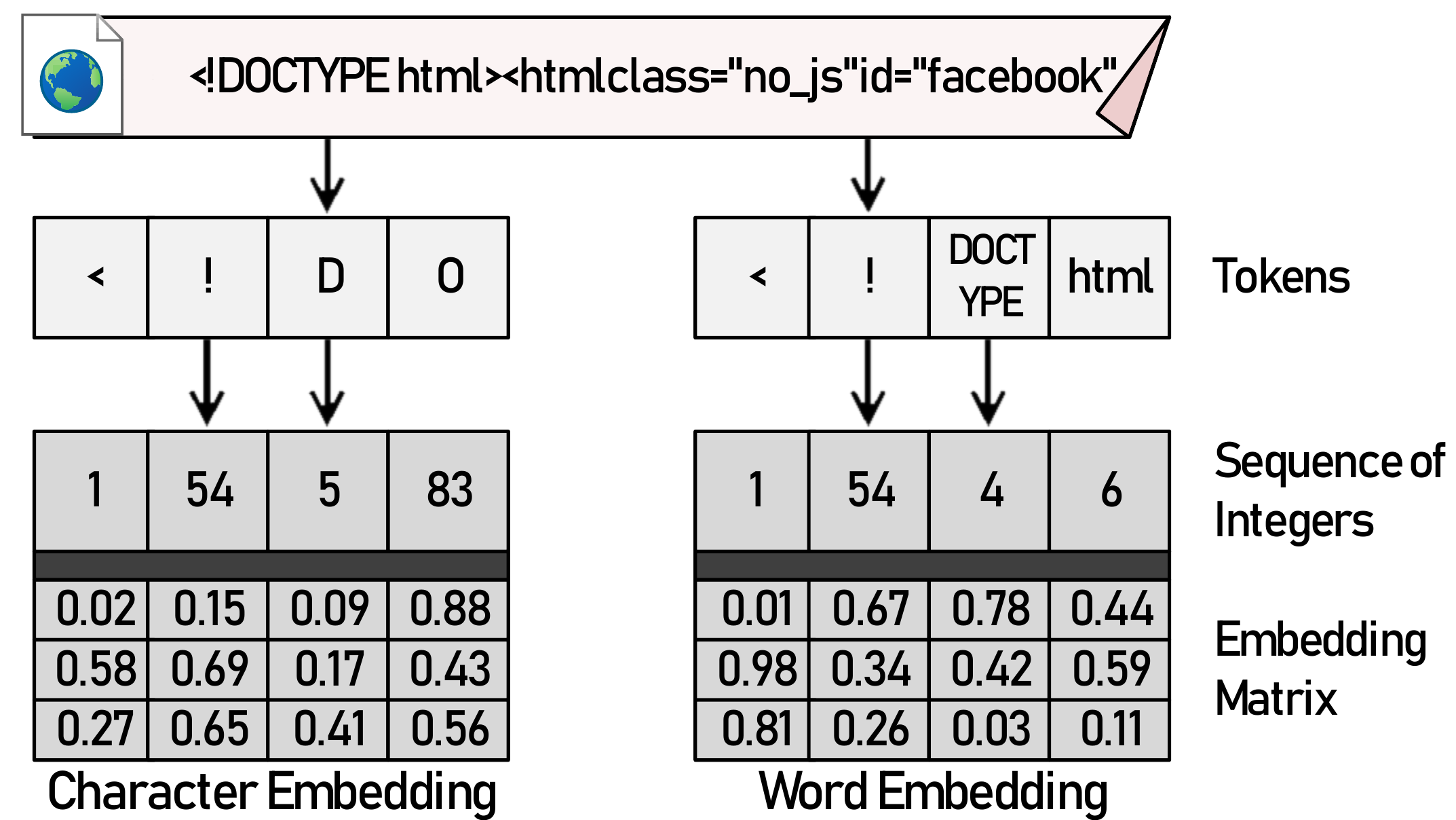}
\caption{Configuration of the Embedding Layer}
\label{embedding}
\end{figure}

To generate the character-level embedding matrix $C_{em}$, the model learns an embedding, which takes the characteristics of the characters in an HTML document. To do so, all the distinct characters, including punctuation marks in the corpus, are listed. We obtained 167 unique characters. We set the length of the sequences ${maxlen_1} = 180$ characters. Every HTML document with strings greater than 180 characters is cut from the 180th character, and any HTML document with characters smaller than 180 characters would be padded up to 180 with zeroes. Before each character in our work is embedded into a d-dimensional vector, we conduct a \textit{tokenization} on the characters in the HTML document and segment the characters into \textit{tokens} as shown in Figure \ref{embedding}. An index is associated with each token before being applied to a d-dimensional character embedding vector where d is set at 100, which is automatically initialised and learnt together with the remainder of the model. To facilitate its implementation, each HTML document \textit{html} is transformed into a matrix, \textit{html} \textrightarrow c $ \epsilon \mathbb{R}^{maxlen_1 \times d} $, where d = 100 and $maxlen_1$ = 180.

For the word embedding matrix $W_{em}$, firstly, the raw HTML document is processed into word-level representations by the word embedding layer. To achieve this, all the different words in the HTML document of the training corpus are listed using the following approach: An HTML document is split into individual words while treating all punctuation characters as separate tokens. For example, as shown in Figure \ref{embedding}, $<! DOCTYPE$ $html$$>$, will be split into $['<', $ $'!', $ $'DOCTYPE', $ $'html'] $. We surmise that punctuation marks provide important information benefits for phishing HTML document detection since punctuation marks are more prevalent and useful in the context of HTML documents than ordinary languages. HTML contains a sequence of markup tags that are used to frame the elements on a website. Tags contain keywords and punctuation marks that define the formatting and display of the content on the Web browser. The listed unique words are used to create a dictionary where every word becomes a feature. We obtained about 321,009 unique words in our dataset. We also padded the HTML documents to make the lengths of the HTML documents uniform in terms of number of words $(maxlen_2 = 2000)$. Each unique word is then embedded into a d-dimensional vector, where d is set at 100, which is automatically initialised and learned together with the remainder of the model. All the HTML documents are converted to their respective matrix representation $(maxlen_2 \times d)$, on which the CNN is applied where d = 100 and $maxlen_2$ = 2000. Figure \ref{embedding} shows an overview of the character and word embedding layer.

We can now introduce Convolutionary layers using the HTML document matrix (for all the HTML documents $s_t \forall t = 1, . . . ,T )$ as the corpus. We applied 32 Convolutionary filters $M \epsilon \mathbb{R}^{d \times n}$ where \textit{n} 8. The Max-Pooling layer whose features are then passed to a 10 unit dense layer comes after the Convolutionary filters. The dense layer, which is regularised by dropout, finally connects to a Sigmoid layer. Then using the ADAM optimisation algorithm \cite{kingma2015adam}, we train the model through backpropagation.

\subsection{Baseline Models}
\begin{figure*}
\centering
\includegraphics[width = 0.85\textwidth, height = 9.5cm ]{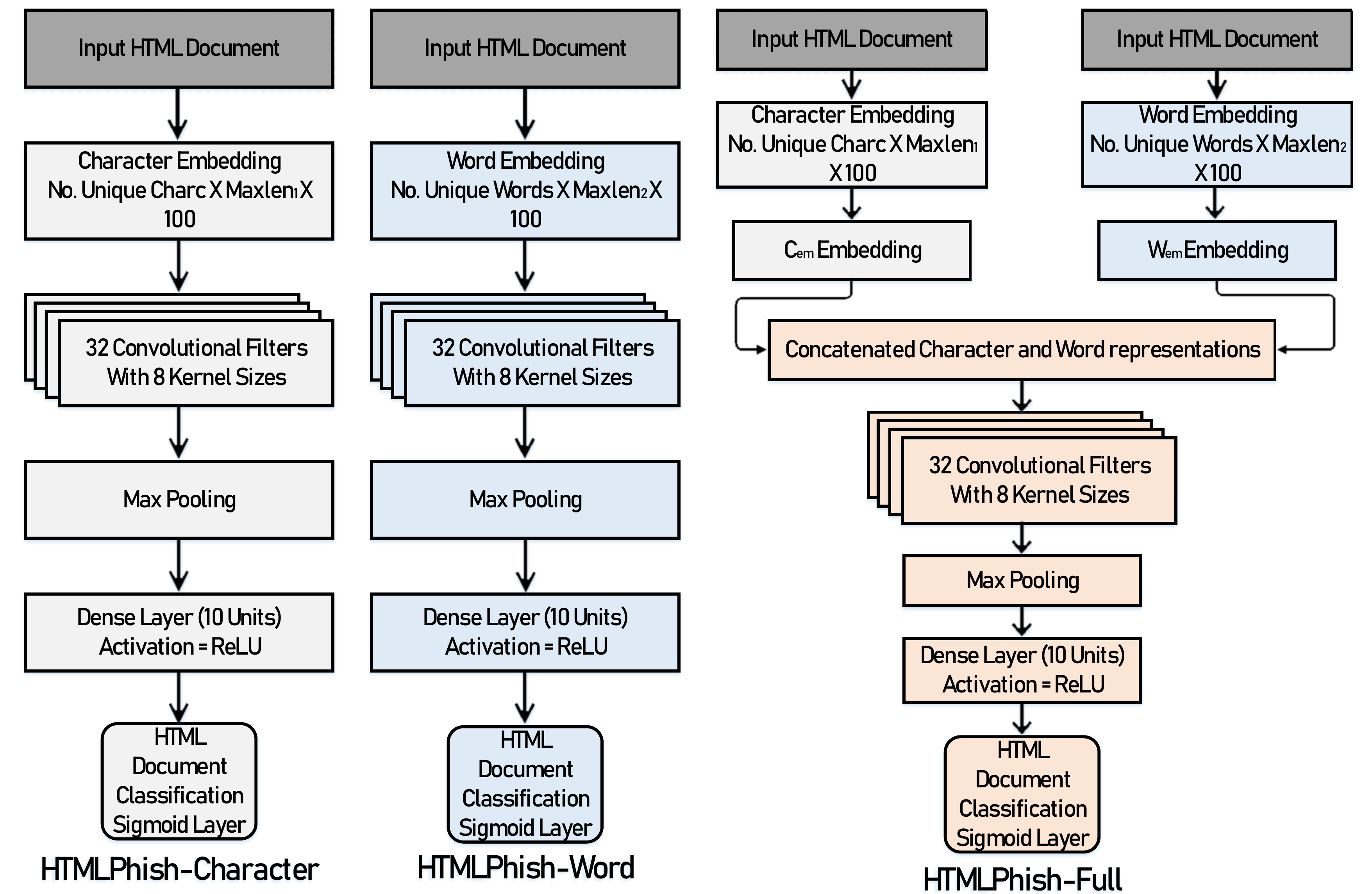}
\caption{The Overall Architecture of HTMLPhish Variants}
\label{fig:HTMLPhish}
\end{figure*}

The baseline models, HTMLPhish-Character and HTMLPhish\hyp{}Word, whose architectures are detailed in Figure \ref{fig:HTMLPhish}, are CNN models trained either on character-level embeddings or word-level embeddings, respectively. The embedding matrices described above are applied to 32 Convolutionary filters $M \epsilon \mathbb{R}^{d \times n}$ where \textit{n} 8. The next layer after the Convolutionary filters is the  Max-Pooling layer, whose features are then passed to a 10 unit dense layer. The Dense layer, which also is regularised by dropout, finally connects to a Sigmoid layer. Also, the models are trained through backpropagation using the ADAM optimisation algorithm.

\begin{table}
 \small\addtolength{\tabcolsep}{-1pt}
\begin{minipage}[t]{\columnwidth}                       
 \begin{center}
  \captionof{table}{HTML Documents Used in this Paper}
  \label{table:2}
      \begin{adjustbox}{width=\columnwidth}
  \begin{tabular}{|p{1.4in}|p{1in}|p{1in}|}
      \hline
      \textbf{Dataset} & \textbf{D1} & \textbf{D2} \\
      \hline
      \hline
      \textbf{Date generated} & 11 - 18 Nov, 2018 & 10 -17 Jan, 2019 \\
      \hline
     \textbf{Legitimate Web Pages} & 23,000 & 24,000 \\
      \hline
      \textbf{Phishing Web pages} & 2,300 & 2,400 \\
      \hline
      \textbf{Total} & 25,300 & 26,400 \\
      \hline

 \end{tabular}

\end{adjustbox}
  \end{center}
\end{minipage}
\end{table}

\begin{figure*}
\centering
\includegraphics[width = 0.85\textwidth, height = 9.0 cm]{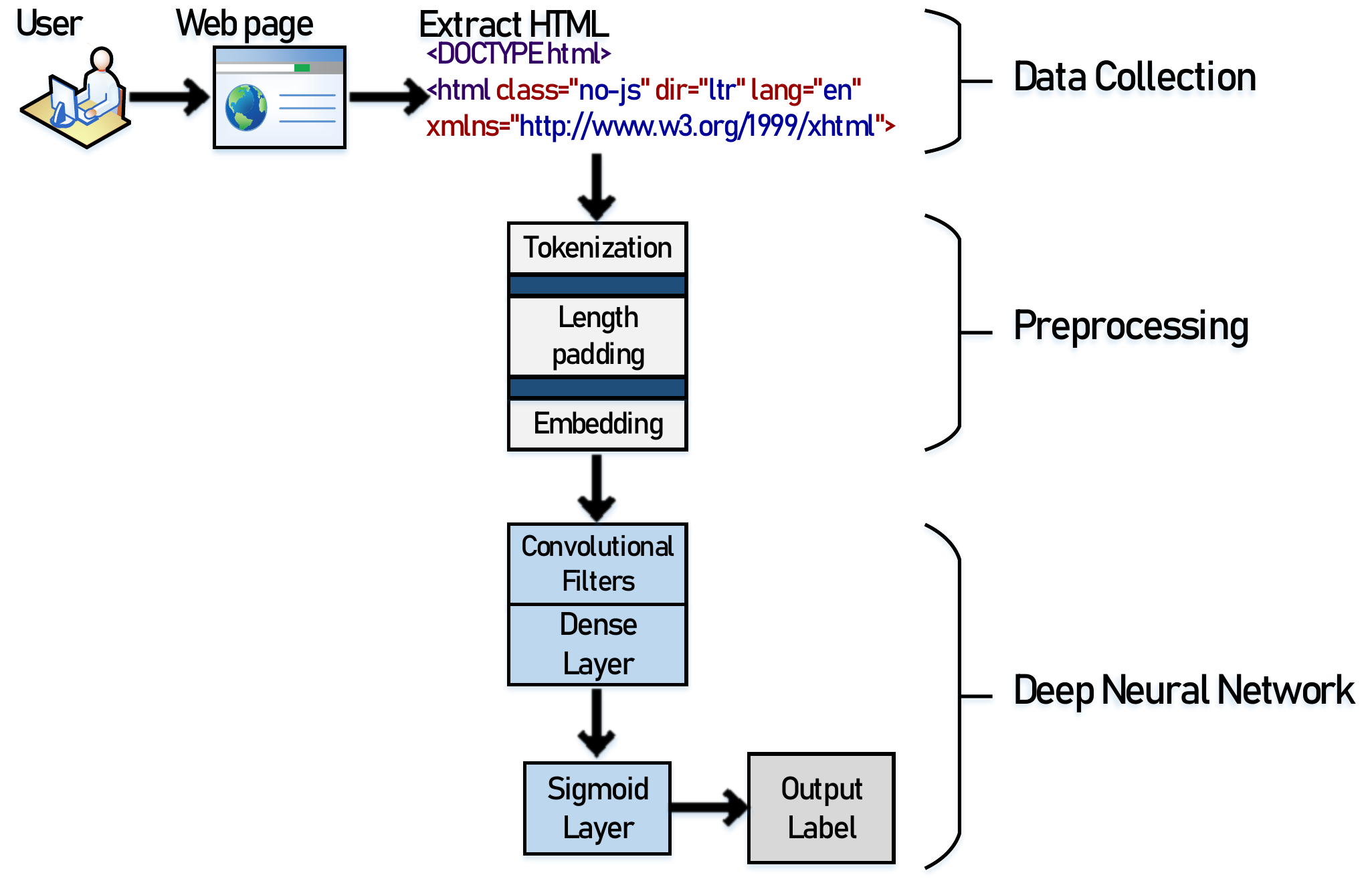}
\caption{A Schematic Overview of the Stages Involved in Our Proposed Model}
\label{model}
\end{figure*}

\section{Dataset}
Data collection plays an essential role in phishing web page detection. In our approach, we collated HTML documents using a web crawler. We used the Beautiful Soup \cite{BeautifulSoup} library in Python to create a parser that dynamically extracted the HTML document from each final landing page. We chose to use Beautiful Soup for the following reasons: 

(1) it has functional versatility and speed in parsing HTML contents, and 

(2) Beautiful Soup does not correct errors when analysing the HTML Document Object Model (DOM). The HTML documents in our corpus include all the contents of an HTML document, such as text, hyperlinks, images, tables, lists, etc. Figure \ref{model} shows an overview of the data
collection stage.

\subsection{Data Collection}
Since phishing campaigns follow temporal trends in the composition of web pages, the earliest data obtained should always be used for training and the most recent data collected for testing \cite{marchal2018designing}. Different phishing pages created during the same time may probably have the same infrastructure. This could exaggerate an over-trained classification model's predictive output. To ensure our evaluation settings reproduces real-world situations in which models are applied on data generated up to the present point and applied on new web pages, we collected a dataset of HTML documents from phishing and legitimate web pages over 60 days.

Also, to ensure the deployability of our model to real-word systems, our data set is required to provide a distribution of phishing to benign web pages obtainable on the Internet in the real-world $(\approx 10/100)$ \cite{whittaker2010large, zhang2007cantina}. Given that when a balanced dataset (1/1), is used, the results can yield a baseline error \cite{borgida1981base}. Consequently, our training dataset \textbf{D1} consisting of HTML documents from 23,000 legitimate URLs and 2,300 phishing URLs was collected between 11 November 2018 to 18 November 2018. \textbf{D1} dataset was used to train and validate the three different variants of our model (HTMLPhish-Character, HTMLPhish-Word, and HTMLPhish-Full). From 10 January 2019 to 17 January 2019, testing data set \textbf{D2} consisting of HTML document from 24,000 legitimate URLs and 2,400 phishing URLs were generated.

Note that $\textbf{D1} \cap \textbf{D2} = \emptyset$. Also, our testing dataset \textbf{D2}, is slightly larger than our training dataset \textbf{D1}. This is because learning with fewer data, and having decent tests on a broader test data means that the detection technique is generalised. This ensures that the features and model of classification include specific features from legitimate and phishing web pages and that the approach can be applied to the vast number of online Web pages. In total, our corpus was made up of 47,000 legitimate HTML documents and 4,700 phishing HTML documents, as shown in Table \ref{table:2}.

The legitimate URLs were drawn from Alexa.com's top 500,000 domains, while the phishing URLs were gathered from continuously monitoring Phishtank.com. The web pages in our dataset were written in different languages. Therefore, this does not limit our model to only detecting English web pages. We manually sanitised our corpus to ensure no replicas or web pages that are pointing to empty content. Alexa.com offers a top list of working websites that internet users frequently visit, so it is an excellent source to be used for our aim.

  
       

\begin{table}
\small\addtolength{\tabcolsep}{-1pt}
\begin{minipage}[t]{\columnwidth}                      
  \begin{center}
  \captionof{table}{HTMLPhish-Full Deep Neural Network}
  
  \label{table:1}
      \begin{adjustbox}{center, width=\columnwidth}
    \begin{tabular}{|p{1.1in}|p{1.2in}|p{0.6in}|}
      \hline
       
    \textbf{Layers} & \textbf{Values} & \textbf{Activation} \\
\hline
\hline
Embedding & Dimension = 100 & - \\
\hline
Convolution & Filter = 32, Filter Size = 8 & ReLU \\
      \hline
Max Pooling & Pool Size = 2 & - \\
      \hline
Dense1 & No. of Neurons = 10, Dropout = 0.5 & ReLU \\
\hline
Dense2 & No. of Neurons = 1 & Sigmoid \\
\hline
Total Number of Trainable Parameters & 412,388,597 & - \\
\hline

\end{tabular}
\end{adjustbox}
  \end{center}
\end{minipage}
\end{table}

\section{Evaluation of HTMLPhish Variants}

\subsection{Experimental Setup}
Table \ref{table:1} details the selected parameters we found gave the best performance on our dataset bearing in mind the unavoidable hardware limitation for our proposed HTMLPhish variants:

a. HTMLPhish-Character

b. HTMLPhish-Word

c. HTMLPhish-Full

The three CNN models were implemented in Python 3.5 on a Tensorflow backend and a learning rate of 0.0015 in the Adam optimizer \cite{kingma2015adam}. The batch size for training and testing the model were adjusted to 20.

All HTMLPhish and baseline experiments were conducted on an HP desktop with Intel(R) Core CPU, Nvidia Quadro P600 GPU, and CUDA 9.0 toolkit installed.

\subsection{Evaluation Metrics}
Because of the severely imbalanced nature of our dataset, we evaluated the performance of our models in terms of the Area under the ROC Curve
(AUC). We also used the receiver operating characteristic (ROC) curve in our evaluation. The ROC curve is a probability curve, while the AUC depicts how much the model can distinguish between two classes, which for our model is - legitimate or phishing. The higher the AUC value, the better the performance of the model. The ROC curve is plotted with the true positive rate (TPR) against the false positive rate (FPR) where $TPR = \frac {(TP)}{(TP+FN)}$ and $FPR = \frac {(FP)}{(TN+FP)}$. Where TP, FP, TN, and FN stand for the numbers of True Positives, False Positives, True Negatives, and False Negatives, respectively.

Additionally, we employed the precision, True Positive Rate, and F-1 score metrics to evaluate the performance of HTMLPhish and the baseline models. The True Positive Rate computes the ratio of phishing HTML documents that are detected by the models. In contrast, the precision metrics compute the ratio of detected phishing HTML documents that are actual phishes to the total number of detected phishing HTML documents.

\begin{table*}
\small\addtolength{\tabcolsep}{-1pt}
\centering
\begin{minipage}[t]{\textwidth}                       
 \begin{center}
  \captionof{table}{Result of HTMLPhish and Baseline Evaluations on the D1 dataset}
  \label{table:3}
     \begin{adjustbox}{center, width=0.9\columnwidth}
  \begin{tabular}{|p{1.2in}| p{0.6in}| p{0.6in}| p{1.1in}| p{0.6in}| p{0.5in}| p{0.8in}| }
   \hline
   \textbf{Models}   & \textbf{Accuracy} & \textbf{Precision} & \textbf{True Positive Rates} & \textbf{F-1 Score} & \textbf{AUC} & \textbf{Training time}\\
      \hline
      \hline
      \textbf{HTMLPhish-Full} &\textbf{0.98} & \textbf{0.97} & \textbf{0.98} & \textbf{0.97} & \textbf{0.93}  & \textbf{6.75 mins} \\
      \hline
      HTMLPhish-Word &0.94 & 0.93 & 0.94 & 0.93 & 0.88 & 10 mins\\
      \hline
      HTMLPhish-Character & 0.95 & 0.92 & 0.95 & 0.94 & 0.90 & 3.5 mins\\
      \hline
      \cite{wei2019deep} & 0.97 & 0.96 & 0.97 & 0.96 & 0.93 & 5.25 mins\\
      \hline
   \cite{bahnsen2017classifying} & 0.95 & 0.94 & 0.95 & 0.94 & 0.91 & 18 mins\\
      \hline
 \end{tabular}
\end{adjustbox}
 \end{center}
\end{minipage}
\end{table*}

\begin{table*}
\small\addtolength{\tabcolsep}{-1pt}
\centering
\begin{minipage}[t]{\textwidth}                       
 \begin{center}
  \captionof{table}{Result of HTMLPhish and Baseline Evaluations on the D2 dataset}
  \label{table:4}
     \begin{adjustbox}{center, width=0.9\columnwidth}
  \begin{tabular}{|p{1.2in}| p{0.6in}| p{0.6in}| p{1.1in}| p{0.6in}| p{0.5in}| p{0.8in}| }
   \hline
   \textbf{Models}   & \textbf{Accuracy} & \textbf{Precision} & \textbf{True Positive Rates} & \textbf{F-1 Score} & \textbf{AUC} & \textbf{Testing time}\\
      \hline
      \hline
      \textbf{HTMLPhish-Full} &\textbf{0.93} & \textbf{0.92} & \textbf{0.93} & \textbf{0.91} & \textbf{0.88}  & \textbf{9 seconds} \\
      \hline
      HTMLPhish-Word &0.90 & 0.87 & 0.91 & 0.88 & 0.73 & 107 seconds\\
      \hline
      HTMLPhish-Character & 0.91 & 0.89 & 0.91 & 0.89 & 0.77 & 7 seconds\\
      \hline
      \cite{wei2019deep} & 0.91 & 0.84 & 0.91 & 0.87 & 0.73 & 15 seconds \\
      \hline
   \cite{bahnsen2017classifying} & 0.90 & 0.90 & 0.92 & 0.90 & 0.78 & 112 seconds\\
      \hline
 \end{tabular}
\end{adjustbox}
 \end{center}
\end{minipage}
\end{table*}

\subsection{Overall Result}\label{Overall Result}
To record the performance of HTMLPhish-Full and the baseline models on the D1 dataset, we split the dataset into 80\% for training, 10\% for validation, and 10\% for testing. Also, taking cognizance of how our data is severely imbalanced, we ensured we manually shuffled the datasets before training. 

\begin{figure}[ht]
  \includegraphics[width=9cm]{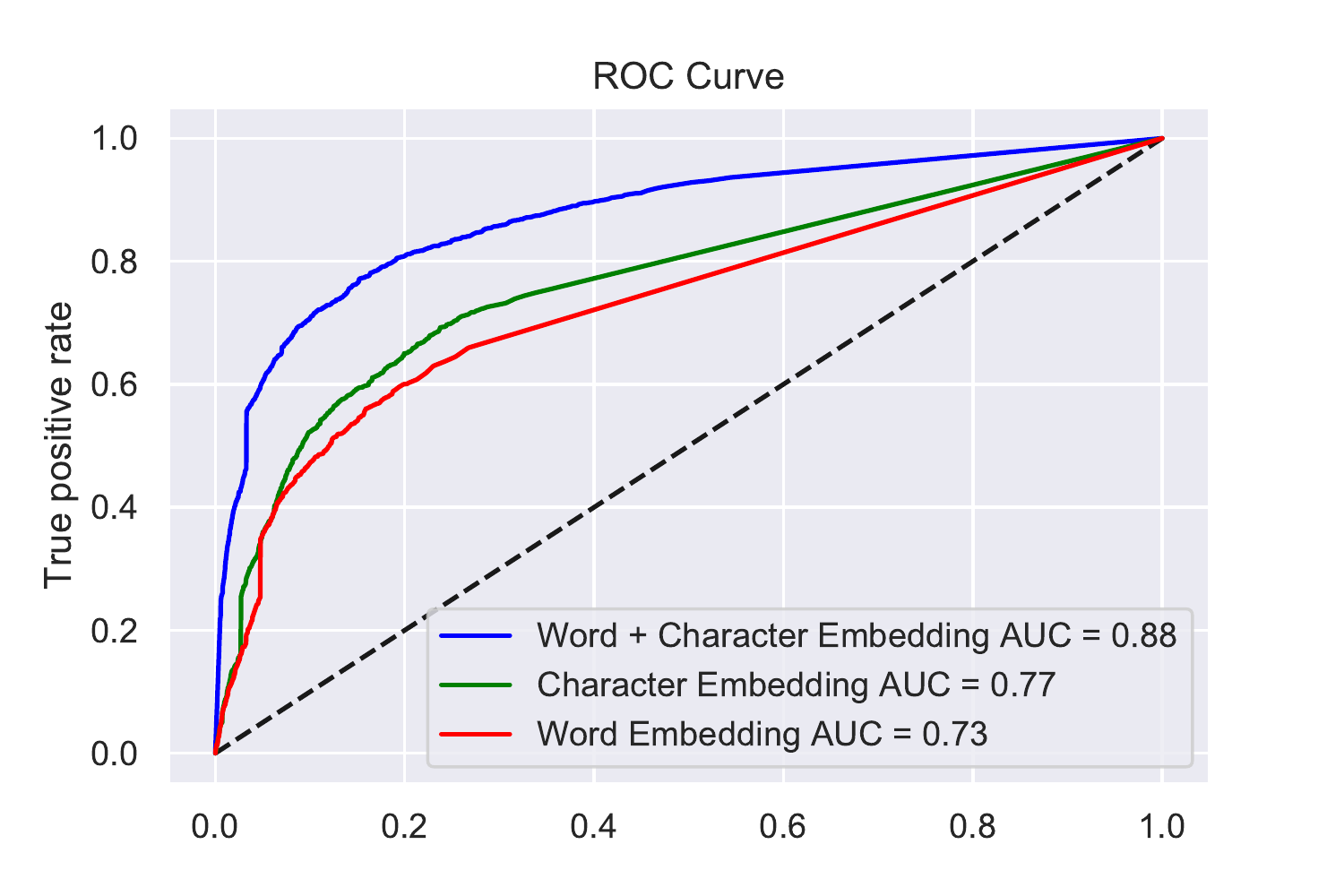}
  \caption{The ROC Curve of HTMLPhish Variants}
  \label{fig: ROC_HTMLPhish + RNN}
\end{figure}

The ROC curves of HTMLPhish and its variants are shown in Figure \ref{fig: ROC_HTMLPhish + RNN}. From the result detailed in Table \ref{table:3}, in general, HTMLPhish-Full significantly outperforms the other two variants: HTMLPhish\hyp{}Character, and HTMLPhish-Word. While HTMLPhish-Character and HTMLPhish-Word have similar performances, HTMLPhish-Full takes advantage of the strengths of both and produces more consistently better results. Also, HTMLPhish-Full offered a significant jump in AUC over the other variants, while HTMLPhish-Word performs slightly worse amongst the three.

On the D1 dataset, HTMLPhish-Full provided a 98\% accuracy and 2\% False Positive Rate. The minimal False Positive Rates indicates the ratio of legitimate web pages, which are incorrectly identified as a phish. This is helpful when the model will be deployed in real-world scenarios as users will not be inappropriately blocked from accessing legitimate web pages.

Considering the computational complexity of HTMLPhish-Full, it can be seen that on a dataset of over 25,000 HTML documents,  HTMLPhish-Full can be speedily trained within 7 minutes. Once trained, HTMLPhish-Full can evaluate an HTML document in 1.4 seconds. 


\subsection{Comparison with State-Of-The-Art Techniques}
We compared HTMLPhish-Full with the methodology, speed, and performance of existing state-of-the-art models in \cite{bahnsen2017classifying} and \cite{wei2019deep}. \cite{wei2019deep} is a Deep Neural Network with multiple layers of CNNs that takes as input word tokens from a URL  to determine the maliciousness of the associated web page. On the other hand, \cite{bahnsen2017classifying} takes as input the character sequence of a URL and models its sequential dependencies using Long short-term memory (LSTM) neural networks to classify a URL as phishing or benign. We applied these techniques to the HTML documents in the D1 dataset and also tested them on the D2 dataset.

From the result detailed in Table \ref{table:3} and Table \ref{table:4}, HTMLPhish-Full provides better precision, recall and comparable accuracy against the existing state-of-the-art models. The performance of HTMLPhish-Word and \cite{wei2019deep} can be attributed to the fact that it is trained on a definite dictionary of words from the training data. Therefore it will be unable to obtain useful embeddings for new words in the test data. HTMLPhish\hyp{}Character and \cite{bahnsen2017classifying} perform better with respect to the AUC metric because the individual character embedding CNN can learn structural patterns in the HTML document and can also obtain feature representations for new words. This makes it easy to be applied to the test data. In addition, due to the limited number of characters, the scale of the CNN model using the individual embedding character remains fixed when compared to word-based model sizes. However, CNN models built with individual character embeddings cannot exploit structural information available in long sequences in the HTML document. It also disregards word borders and makes it challenging to differentiate special characters in the data.

Furthermore, CNN's using only character level embedding struggles to differentiate information for scenarios where phishing HTML documents try to imitate benign HTML documents through small modifications to one or few words in the HTML document\cite{chu2013protect}. This is because the Convolutional filters will likely yield similar output from a sequence of characters with a similar spelling. Therefore, CNNs using only character embeddings are not enough to obtain structural information from the HTML document in detail. That is the reason word embeddings must be taken into account. Consequently, HTMLPhish-Full takes advantage of both word and character embedding matrices to accommodate unseen words in the test data, and therefore yield a better result than the other variants and baseline models.

\subsection{Temporal Resilience}
The techniques for implementing a phishing web page is continuously evolving due to emerging technology applications for designing phishing web pages. The evaluation of the resilience of this evolution is paramount for a phishing web page detection technique. In this paper, we applied the longitudinal study \cite{marchal2017off} by evaluating the accuracy of the HTMLPhish-Full using freshly collected data. This study enabled us to infer a maximum retraining period, for which the accuracy of the system does not reduce. For a security supplier deploying HTMLPhish-Full in the wild, the retraining time frame can provide an approximate cost of maintenance.

Using the evaluation metrics detailed above, we compared the accuracy of HTMLPhish variants and baseline models on the training data \textbf{D1} with its accuracy when applied to the test data \textbf{D2} without retraining the model. From the results in Table \ref{table:4}, HTMLPhish-Full provided a 98\% accuracy on the training dataset while yielding a 93\% accuracy on the test dataset. The result of our longitudinal study demonstrates the readiness of HTMLPhish-Full for real-world deployment. HTMLPhish-Full will remain temporally robust, and will not need retraining within at least two months.

\section{Conclusion}
In this paper, we proposed HTMLPhish, a deep learning based data-driven end-to-end automatic phishing web page classification approach. HTMLPhish receives the HTML content of a web page as input and applies CNNs to learn the semantic dependencies in both the characters and words in the HTML document in a jointly optimized network. Furthermore, we applied convolutions on a concatenation of the matrix of character and word embeddings in order to ensure the effective embedding of new words in the test HTML documents. Our approach can learn context features from HTML documents without requiring extensive manual feature engineering.

We evaluated our model using a comprehensive dataset of HTML contents presented in a real-world distribution. HTMLPhish provided a high precision rate, showing a temporally stable result even when it was trained two months before being applied to a test dataset.

The future work is to compare our model to feature engineering-based models that extract features only from the HTML document. Also, we intend to implement our model as a browser extension. This will enable HTMLPhish to recognise phishing websites in real-time. 

\section*{Acknowledgment}{The authors acknowledge the Petroleum Technology
Development Fund (PTDF), Nigeria for the funding and
support provided for this work.}

\medskip
\bibliographystyle{IEEEtran}
\bibliography{main}

\end{document}